\documentclass[fleqn,usenatbib]{mnras}

\usepackage{newtxtext,newtxmath}

\usepackage[T1]{fontenc}
\DeclareRobustCommand{\VAN}[3]{#2}
\let\VANthebibliography\thebibliography
\def\thebibliography{\DeclareRobustCommand{\VAN}[3]{##3}\VANthebibliography}

\usepackage[utf8]{inputenc}
\usepackage{graphicx}	

\usepackage{amsmath}	
\usepackage{url}

\usepackage{ulem}
\usepackage{multicol}
\def\bea{\begin{eqnarray}}
\def\ena{\end{eqnarray}}

\begin{document}

\title[Low-frequency observations of giant pulses from ordinary pulsars]{Low-frequency observations of giant pulses from ordinary pulsars}

\author [A.~N.~Kazantsev and M.~Yu.~Basalaeva] 
{
A.~N.~Kazantsev$^{1}$\thanks{E-mail:kaz.prao@bk.ru (ANK)}, 
~M.~Yu.~Basalaeva$^{2}$, 
\\
$^{1}$P. N. Lebedev Physical Institute of the Russian Academy of Sciences,\\ Pushchino Radio Astronomy Observatory,
Pushchino 142290, Russia
\\
$^{2}$ Astrophysical school Traektoria,\\  107078, Moscow, Russia
}

\date{}


\maketitle

\label{firstpage}

\begin{abstract}
We present results of the investigation of the giant radio pulses (GRPs) generation rate from 5 radio pulsars (B0301+19, B0950+08, B1112+50, B1133+16, and B1237+25) and anomalous intensity pulses generation rate from B0809+74. All data used were obtained with Large Phased Array radio telescope of Pushchino Radio Astronomy Observatory at 111~MHz from 2012 to 2021. In addition to analysis of the rate of generation of bright pulses, we analyze the distribution of bright pulses in the phase of the pulsar period and search for clusters of bright pulses - several bright pulses emit in adjacent pulsar periods. It is found that pulsars B0301+19, B1112+50, B1133+16, and B1237+25 demonstrate different generation rates and generation of clusters. Pulsar B1112+50 more often than others studied pulsars generates GRPs cluster. The longest cluster of GRPs containing four single pulses is detected from this pulsar. GRPs from studied pulsars are distributed along the longitudes of the main components of average pulses of these pulsars. This distribution is 1.5-2 times narrower than the phase distribution of non-giant pulses. It is found that the distance between the components of the average GRP profile and the distance between the components of the average non-giant profile differ substantially for pulsars with multicomponent average profiles.
\end{abstract}



\section{Introduction}

The large group of radio pulsars shows small pulse-to-pulse intensity variations within the limits of 10 times of the intensity of the averaged pulse. However, some pulsars demonstrate unpredictable short-duration outbursts of pulsed radio emission. These pulses were named Giant Radio Pulses (GRPs).

At the present day, only about 16 pulsars are known to be GRP emitters \citep{Staelin1968, Wolszczan1984, Singal2001, Ershov2003, Johnston2003, Joshi2004, Knight2005, Kuzmin2004, Ershov2005, Kuzmin2006, Crawford2013, Kazantsev2017b, Kazantsev2019}, i.e.  emit individual pulses that satisfy GRP criteria (the peak flux densities of GRPs exceed the flux density in the average profile of pulsar by a factor of 30). These pulsars can be divided into two groups: pulsars with strong  magnetic fields at their light cylinders (over $B_{LC} > 10^{5}$ Gauss) and millisecond and ordinary pulsars with $B_{LC}$ from several to several hundred Gauss. Well-known representatives  of the  former group are the Crab pulsar \citep{Staelin1968} and millisecond pulsar B1937+21 \citep{Wolszczan1984}. There are a lot of theoretical models, concerning GRPs from bright, fast-rotating pulsars of the first group \citep{Hankins2003, Istomin2004, Petrova2006a, Machabeli2019}, but much less is known about pulses from the ordinary pulsars with GRPs. 

Still, we consider that studies of GRPs from the radio pulsars are equally important for understanding the pulse generation mechanism. In early work \citep{Kazantsev2018}, we have analyzed the energy characteristics of this group of pulsars. However, we have not studied temporal  properties  of GRP generation. This issue is  clearly very important, because it can shed light  on the  dynamical behavior of GRPs generation mechanism.

Here we present a statistical study of the rate of GRP emission from six pulsars: B0301+19, B0809+74, B0950+08, B1112+50, B1133+16 and B1237+25. All the listed pulsars were previously noted as the generators of sporadic bright individual pulses, mainly at low frequencies. A detailed study of the GRPs of the pulsar B0301+19 has been carried out in \cite{Kazantsev2019}. In this investigation, an analysis of 257 observation sessions at 111 MHz has been performed, and it has been shown that 2.5\% of individual pulses of the pulsar exceed the peak flux density in the average profile by a factor of 30. The distribution of individual pulses of this pulsar has a complex shape, with a pronounced power-law tail formed by bright individual pulses. \cite{Ulyanov2006} has found very bright pulses 50-100 times higher than the average profile in amplitude from the pulsar B0809+74. The authors call these pulses "anomalous intensity pulses", for this reason we use this definition further to describe bright pulses from the pulsar B0809+74. The emission of pulses of anomalous intensity can be similar to the GRPs emission, and the long-term analysis of these pulses can be very useful for understanding the nature of their generation. Giant pulses from the pulsar B0950+08 have been detected at 103 MHz in \cite{Singal2001}. The authors have analyzed about 1 million individual pulsar pulses received with the Rajkot radio telescope and shown that about 1 percent of these pulses exceed the average pulsar profile by more than 100 times. The analysis of the distribution of pulses by the peak flux density, however, has not carried out in that investigation. Later, in \cite{cairns2004}, it has been discovered that there are several components of radiation from B0950+08: GRPs emission, emission of giant micropulses, and several others. All these components have a power-law of energy distribution and differ in power-law indices. Lack of similar investigations for other pulsars with GRPs, such as B0031-07, B0301+19, B1133+16, etc. does not allow us to assert unambiguously that these emissions are characteristic only of the pulsar B0950+08. At the same time, it is impossible to unambiguously distinguish giant pulse and several merged giant micropulses from each other in observations of pulsars with emission similar to that of B0950+08  at low frequencies and with insufficient time resolution. Such an analysis requires observations at high frequencies, as has been done in \cite{cairns2004}. The analysis of powerful individual pulses from B0950+08 has been carried out at low frequencies in \cite{singal2012} and \cite{Smirnova2012} and has been shown the aggregate statistics of these emissions unambiguously illustrates this pulsar as a pulsar with GRPs, which makes it possible to include it in our sample. GRPs from the pulsar B1112+50 have been discovered in \cite{Ershov2003}. The authors have analyzed 172 observation sessions and found 126 GRPs (on average 1 GRP per 150 pulsar periods), exceeding the flux in the average profile by a factor of 30. The authors also noted the generation of GRPs from this pulsar in pairs and triplets, and estimated the period of pulsar activity at several seconds. The authors pointed out the presence of a pronounced power-law tail in the distribution of pulses over the peak flux density, which corresponds to bright individual pulses of the pulsar. \cite{Kramer2003} has reported bright pulses from B1133+16 at 5 GHz. In \cite{Karuppusamy2010} it has been shown that the pulsar B1133+16 emits bright pulses in the range 110-180 MHz. These pulses are 10 times higher than the average pulsar profile in amplitude, and the distribution of individual pulses over peak flux densities is power-law. Similar results have been obtained in this work for the pulsar B1112+50, which is also included in our sample. A more detailed analysis of GRPs from B1237+25 has been made in \cite{Kazantsev2017b}. The pulsar regularly emits pulses with a high flux density in comparison with regular pulses, constant localization at the longitude of one of the components of the average profile, which is from 50 to 100\% of the duration of the corresponding component. The distribution of individual pulses over the peak flux density is well approximated by a power law.

It is important to note once again that not all the above listed pulsars are noted in the literature as sources of GRPs. In addition, several of these pulsars have their own peculiarities in generating bright pulses (for example, B0950+08). Analysis of flare activity at long times for a given sample of active pulsars can provide the necessary material for searching for similarities and differences in the mechanisms of emission of bright individual pulses of these pulsars.

The article has the following structure: in Section~\ref{sec:obs_and_proc} we describe the process of obtaining data and methods of data analysis, Section~\ref{sec:res} contains the results of the analysis for each pulsar. Discussions and conclusions are presented in Section~\ref{sec:dis} and Section~\ref{sec:conclusions}, respectively.

\begin{table*}
\caption{Parameters of pulsars and durations of the observations of each pulsar.} 
\label{tab:pulsar_info} 
\begin{tabular}{ | l | l | l | c | c | c | c | c | c | c |}
\hline
Pulsar name & RA & DEC & P & dP/dt & DM & PEpoch & Years & Sessions & Total observational time\\ 
(epoch 1950) & (J2000) &  (J2000) & (s) & (10$^{-15}$ s/s) & (cm$^{-3}$ pc) & (MJD) & & & (h) \\
\hline
B0301+19 & 03h04m33.115s & $19^{\circ}32'51.4''$ & 1.3876 & 1.29 & 15.66 &  49289.00 & 2012-2021 & 752 & 42.32 \\
B0809+74 & 08h14m59.50s & $74^{\circ}29'05.70''$ & 1.2922 & 0.17 & 5.75 &  49162.00 & 2012-2020 & 534 & 106.03 \\
B0950+08 & 09h53m9.3097s & $07^{\circ}55'35.75''$ & 0.2531 & 0.23 & 2.97 & 46375.00 & 2012-2021 & 1179 & 63.4 \\
B1112+50 & 11h15m38.400s & $50^{\circ}30'12.29''$ & 1.6564 & 2.49 & 9.19 & 49334.00 & 2012-2020 & 1540 & 128.25 \\
B1133+16 & 11h36m3.1198s & $15^{\circ}51'14.183''$ & 1.1879 & 3.73 & 4.84 & 46407.00 & 2013-2020 & 1050 & 56.11 \\
B1237+25 & 12h39m40.4614s & $24^{\circ}53'49.29''$ & 1.3824 & 0.10 & 9.25 & 46531.00 & 2012-2021 & 781 & 45.77 \\
\hline

\end{tabular}
\end{table*}

\section{Observations and data processing}
\label{sec:obs_and_proc}

Our observations were made with the Large Phased Array transit radio telescope of Pushchino Radio Astronomy Observatory (LPA LPI) at the Astro Space Center, P.N.Lebedev Physical Institute. This is a low-frequency radio telescope with a central frequency is equal to 111~MHz, and effective bandwidth is equal to 2.3~MHz. During our observational sessions, the telescope effective area was 20~000~$\pm$~1300~$m^2$ in the zenith direction. The single linear polarization was used.We used the sampling interval 1.2288~ms for all observational sessions. The LPA LPI observes pulsars during their culmination only, when they cross the beam of the LPA LPI. As a result, the duration of observations depends on the declination of a pulsar. For example, the duration of a single observational session of B0809+74 is equal to $\sim$11~minutes, and B1237+25 - $\sim$3.5~minutes. 

Our data processing pipeline can be described as follows. The digital receiver gets the phased analog signal from the radio telescope. After that, using the synchronizer that generates trigger pulses, the signal is split into a sequence of segments with the duration is equal to a period of a given pulsar. The synchronizer has $\pm$100 ns accuracy when converting GPS time to the universal time scale and $\pm$10 ns accuracy when setting the start time of the receiver. These levels of accuracy are sufficient for observing normal-period pulsars at the central frequency of LPA LPI and performing phase analysis of individual pulses from these pulsars.

\begin{figure*}
\begin{multicols}{2}
\begin{center}
\includegraphics[width=2\columnwidth]{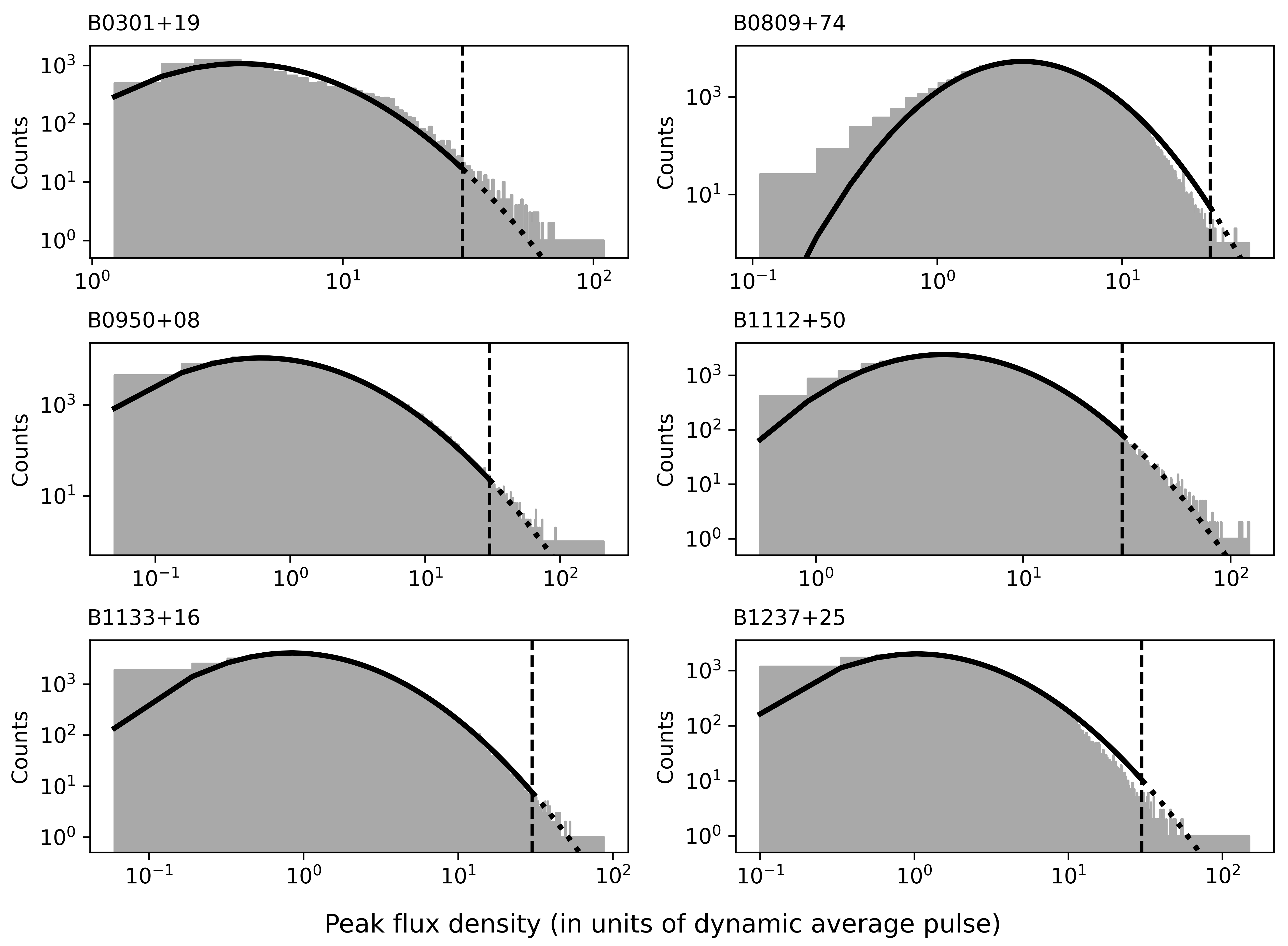}
\end{center}
\end{multicols}
\caption{Distribution of individual pulses by peak flux density in units of the dynamic average profile for each of the studied pulsars. The solid and dotted lines show the fitting of the distribution by the lognormal function. The vertical dash line shows the cutoff level of 30 dynamic average profiles of the pulsar.}
\label{fig:distridutions}
\end{figure*}

The signal is digitized at a frequency of 5 MHz for each pulse. The digitized pulse is accumulated in the receiver’s buffer.
After digitization, all readings for a given pulsar are loaded into the fast Fourier transform hardware processor from the buffer. The processor returns the digitized signal as a raw-file observation. The file consists of a header with common information (name of pulsar, start time of observation, time sample, etc.) and an array of time-series spectra. Data of time-series spectra are recorded as 32-bit floating-point numbers. This algorithm does not include absolute calibration of intensity, and units of intensity are analog-to-digital converter (ADC) units.
The next stage of data processing is the offline de-dispersion procedure \citep{Hankins1975}, which is done at a fixed dispersion measure (DM) for a given pulsar (see Tab.~\ref{tab:pulsar_info}). The de-dispersed pulses for a given observation are saved for a subsequent analysis. The de-dispersed individual pulses of a pulsar recorded in an observing session were summed up synchronously with the pulsar period to form a dynamic average pulse (average pulse per one observing session). The individual pulse is considered to be a GRP if its peak flux density exceeds of that of the dynamic average pulse by 30 times or more. We have used a similar criterion for identifying GRPs in our early studies of these pulsars \citep[e.g.,][]{Kazantsev2017b, Kazantsev2018, Kazantsev2019}. This makes it possible to use the GRPs statics obtained in this and the above-mentioned works in further investigations.  Selection of GRPs in amplitude relative to the average profile of a pulsar has been used in the studies of other authors \citep[e.g.,][]{Ershov2003, Kuzmin2004, Smirnova2012, singal2012, tsai2015}. However, for an extraction of radio giant pulses also the intensity distribution of individual pulses needs to be taken into account since they have been observed to be power-law distributed \citep{Staelin1968}. The corresponding distributions of individual pulses for the studied sample are given in~Fig.~\ref{fig:distridutions}. The Freedman–Diaconis rule \citep{Freedman1981} was used to define the width of the bins that were used in the distributions. A non-linear least squares method was used to fit the distributions. It is important to note that the using of the dynamic average profile as a giant / non-giant separator provides information on the modulation of radiation within a single observation session. 

The next parameters were calculated for each detected GRP:

\begin{itemize}
  \item Time stamp of a first sample (MJD),
  \item Time of arrival (MJD),
  \item Amplitude of GRP (ADC units),
  \item STD noise (ADC units),
  \item Width of pulse at 50\% of peak (ms),
  \item Width of pulse at 10\% of peak (ms).
\end{itemize}

The rate of GRPs generation were calculated in 30 days bins and normalized to the total duration of observations in corresponding bins. Taking into account the used time scale of the nine years observations, this bin size is optimal, since it allows us to detect variations on scales from months to years. Additionally, a search for clusters of GRPs is performed. A cluster means a sequence of giant pulses emitted in adjacent periods.

A method of control charts, also known as Shewhart charts \citep{Shewhart1930},  were chosen to determine the stability of the rate of GRPs, namely, individuals and moving charts.
The individual chart displays the individual measured values.  The moving range chart displays the difference from one point to the next one. 
These two charts show changes in processes or phenomena that change the mean or variance of the measured statistic. The statistical criterion of instability in these charts is the excess of the mean $\bar{x}$ by three standard deviations $\sigma$. We called a level $\bar{x}+3\sigma$ is an upper control limit (UCL), and a level $\bar{x}-3\sigma$ as lower control limit (LCL).

The value for each point of the individual chart is simply the measurement value $x_i$.
Each moving range is calculated as:
\begin{equation}
MR_i = |x_i - x_{i-1}|,
\label{MR:ref}
\end{equation}
which is the absolute value of each value minus the previous value.
Individuals and moving range without emission above UCL means that GRPs rate is stable. 

The phase analysis was carried out in several stages. The phase distributions of giant and non-giant pulses by the longitude of the averaged pulse were analyzed at the first stage. The phase was detected simply as a maximum point of an individual pulse. During the second stage, the averaged GRP and the non-giant averaged pulse were calculated by summing up all corresponding pulses.
The average pulses of studied pulsars were fitted by combination of Gaussian functions for estimation of a location, a half-width and amplitude of the average pulses.
The following function was used for fitting one component of the average pulse profile:

\begin{equation}
f(t) = ae^{-\frac{(t-b)^{2}}{2\sigma^{2}}},
\label{gauss_formula:ref}
\end{equation}
where:
$a$ is the height of the curve's peak, $b$ is the position of the center of the peak and $\sigma$ is the standard deviation. The double- or three-component average pulses were fitted by the sum of two or three Gaussian functions, respectively.

%

\section{Results}
\label{sec:res} 

\begin{table}
\caption{XmR chart parameters} 
\label{tab:rate_info} 
\begin{tabular}{ | l | c | c | c | c | c | c | }
\hline
Parameters & B0301 & B0809 & B0950 & B1112 & B1133 & B1237\\
\hline
Mean(x) &  0.12 &  0.01  &  0.44 & 0.26 & 0.07 & 0.06 \\
UCL(x)  & 0.45  &  0.03 &  1.18 & 0.47 & 0.28 & 0.33 \\
LCL(x)  & -0.21 &  -0.02 & -0.31 & 0.05 & -0.14 & -0.20 \\
Mean(MR) & 0.09 &  0.01  & 0.20 & 0.08 & 0.06 & 0.07 \\
UCL(MR)  & 0.34 &  0.03  & 0.77 & 0.28 & 0.23 & 0.33 \\
\hline
\multicolumn{7}{l}{Mean(x) is measured process center.} \\
\multicolumn{7}{l}{UCL(x) is +3$\sigma$ upper control limit.} \\
\multicolumn{7}{l}{LCL(x) is -3$\sigma$ lower control limit.} \\
\multicolumn{7}{l}{Mean(MR) is measured moving range center.} \\
\multicolumn{7}{l}{UCL(MR) is is +3$\sigma$ upper control limit of moving range.} \\
\end{tabular}
\end{table}







\begin{table}
\caption{Clusters of GRPs. Numbers of detected pairs of GRPs are in the second column, number of instances of clusters with >2 GRPs are in the third column, and  in the  last column  we show  the maximal size of the clusters for each pulsar.} 
\label{tab:group_info} 
\begin{tabular}{ | l | c | c | c | }
\hline
Name & 2 GRPs & >2 GRPs & Max size of cluster\\ \hline
B0301+19 & 6 & 1 & 3 \\
B0809+74 & 1 & 0 & 2 \\
B0950+08 & 50 & 6 & 3 \\
B1112+50 & 75 & 5 & 4 \\
B1133+16 & 1 & 0 & 2 \\
B1237+25 & 1 & 0 & 2 \\ 
\hline

\end{tabular}
\end{table}

Resultant statistical properties of the rate of GRPs generation is presented in Tab.~\ref{tab:rate_info}. Also, we separately present information about clustered GRP in  Tab.~\ref{tab:group_info}. 
Below, detailed information about each pulsar is provided.

\begin{figure*}
\begin{multicols}{2}
\includegraphics[width=2\columnwidth]{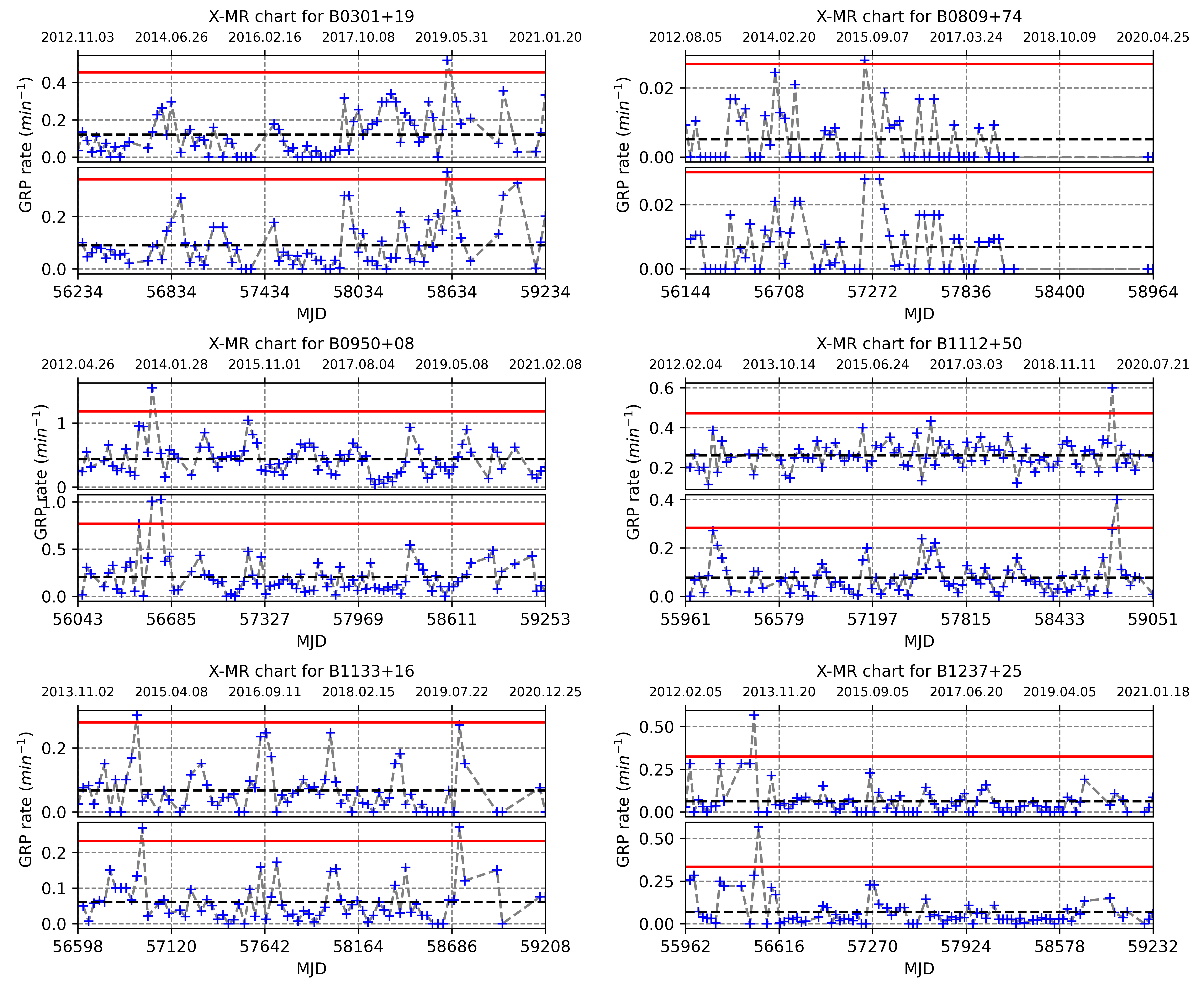}
\end{multicols}
\caption{Individuals (upper panels) and moving range (lower panels) charts of GRPs of studied pulsars. Each point contains data averaged over 30 days.}
\label{fig:rate_gp}
\end{figure*} 

\subsection{PSR B0301+19}

During the processing of 752 observation sessions of the pulsar B0301+19, 14~588 individual pulses were detected with a signal-to-noise ratio (SNR) of more than four out of these, 322 individual pulses were GRPs identified as according to our selection criterion. This pulsar demonstrated quite unstable rate of GRPs generation, since strong modulation is noticeable (see Fig.~\ref{fig:rate_gp}, B0301+19). The average rate is about eight giant pulses per hour. There is a jump in the rate of GRP generation around the MJD~58634 epoch. The maximal rate reaches a value of 31 GRPs per hour. For a significant number of sessions, no individual pulses which could be classified as GRPs were detected. 

\begin{figure*}
\begin{multicols}{2}
\begin{center}
\includegraphics[width=2\columnwidth]{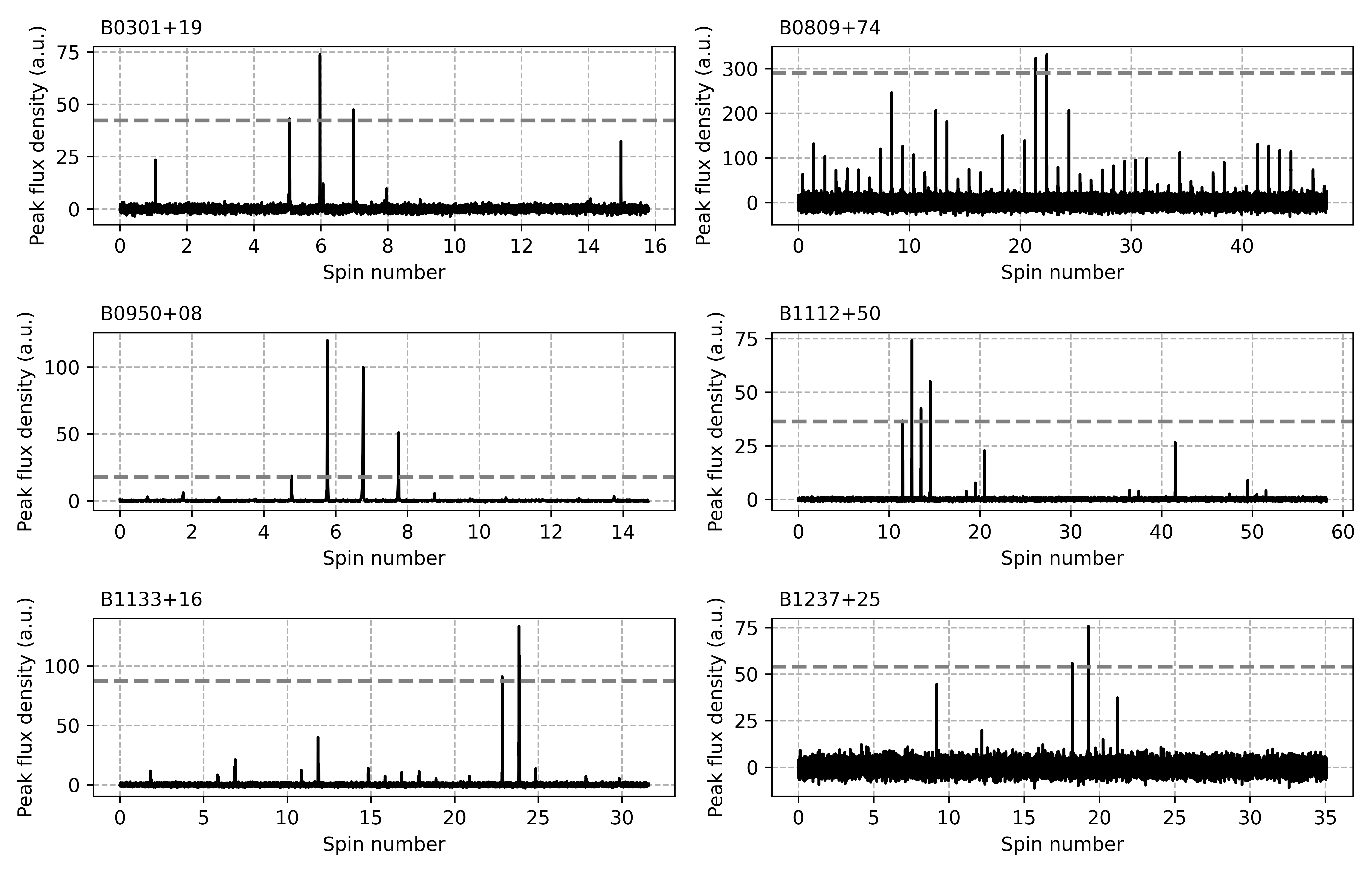}
\end{center}
\end{multicols}
\caption{Examples of GRPs cluster. The gray dashed line shows the GRPs threshold -- i.e. flux density is 30 times larger than the peak flux density of the dynamic averaged pulse.}
\label{fig:group_gp}
\end{figure*}

In the framework of a search for  GRPs clusters, six clusters from B0301+19 were detected. The longest cluster includes three successively emitted GRPs of the pulsar (Fig.~\ref{fig:group_gp}).

The phase distribution of individual pulses from B0301+19 is shown in Fig.~\ref{fig:phases}. The giant pulses of the pulsar are located along the longitudes of the main components of the average pulse out of all the registered GRPs, 124 are localized on the first component of the average profile, and 192 are localized on the second component. 

The fitting of the averaged GRP and average pulse, which includes all registered non-giant pulses, are shown in Fig.~\ref{fig:fitting}. According to the fitting results, the distance between components of the average pulse is (66.1~$\pm$~0.2)~ms. The same parameter for averaged GRP is equal to (64.6~$\pm$~0.2)~ms. In units of the pulsar period, these distances are 0.0477 and 0.0465, respectively.

\subsection{PSR B0809+74}

In 534 conducted observations of PSB B0809+74, 261~108 pulses with SNR > 4$\sigma$ were detected. As we have already mentioned, this pulsar is usually not considered to be a GRPs emitter. Nevertheless, it sometimes generates individual pulses that exceed the amplitude of the average profile by 30 or more times \citep{Ulyanov2006}. We detected 39 such pulses. The rate of anomalous intensity pulse generation was the lowest among the studied pulsars by absolute value (see  Tab.~\ref{tab:rate_info}). There were no strong individual pulses in most observation sessions (see Fig.~\ref{fig:rate_gp}, B0809+74). Nevertheless, there were sessions with an increased generation rate. On average, B0809+74 generates approximately one anomalous intensity pulse in 2 hours of observations. The maximum detected rate of GRPs generation was two pulses per hour.   

We found only one cluster of two consecutive strong pulses (see Fig.~\ref{fig:group_gp}).

Strong pulses are located in the middle of the longitude of the average pulse (see left panel of Fig.~\ref{fig:phases}). In a view of a small number of detected strong pulses, the average pulse is highly jagged (see the right panel of Fig.~\ref{fig:phases}). 

\subsection{PSR B0950+08}

We analyzed 313~227 individual pulses from B0950+08 whose peak flux exceeds 4$\sigma$. Out of these, 1688 were GRPs. Pulsar B0950+08 showed a sinusoidal variation in GRPs generation rate with a period of about 300 days (see Fig.~\ref{fig:rate_gp}). There were only several  'empty bins', where no GRPs were registered. The rate of generation varies between 0.1~$min^ {-1} $ and 1~$min^ {-1} $ with the average value equals to 0.44~$min^ {-1} $or 26 GRPs per hour. There is a jump in rate around epoch MJD 56500 where it increased approximately by 3.5 times, up to 93 GRP per hour.

We detected 50 clusters of GRPs from B0950+08. Six of these include three consecutive GRPs. The example of one of these groups is presented on Fig.~\ref{fig:group_gp}.

The average pulse of pulsar B0950+08 is fairly complex (see the right panel of Fig.~\ref{fig:phases}). It includes a precursor area (0 - 0.12 pulse phase) and a main pulse (0.12 - 0.25 pulse phase). There are 92\% of GRPs on the main pulse longitudes. Most frequently, GRPs were generated at the phase range of the second component of the main pulse of pulsar. As a result, the shapes of averaged GRP and average pulse are quite different. In addition, one can notice an activity of GRPs in the precursor area. It is clearly seen that near MJD 56430 and MJD 57232 the GRPs were actively emitted both in the precursor area and at the trailing edge of the main profile. Strong modulation activity at precursor longitudes have been noted earlier \citep{cairns2004, Smirnova2012}.

\subsection{PSR B1112+50}

In 1540 observation sessions of B1112+50, 65~034 individual pulses were found with a signal-to-noise ratio of more than $4\sigma$ these, 2023 met the GRP criterion. The pulsar B1112+50 demonstrates a quite enough stable rate of GRPs generation (see Fig.~\ref{fig:rate_gp}). Nevertheless, a jump in the rate of generation can be seen around MJD 58740. At a maximum, the rate of GRPs generation reaches 36 pulses per hour.
Compared to the average, this gives about 16 giant pulses per hour.

Also, we detected the largest number of clusters from this pulsar -- 75 (see Tab.~\ref{tab:group_info}). In five cases, the number of GRPs exceeds two and the largest cluster contained four pulses. In  Fig.~\ref{fig:group_gp} we show one of the largest clusters.

\begin{figure*}
\begin{multicols}{2}
\begin{center}
\includegraphics[width=2\columnwidth]{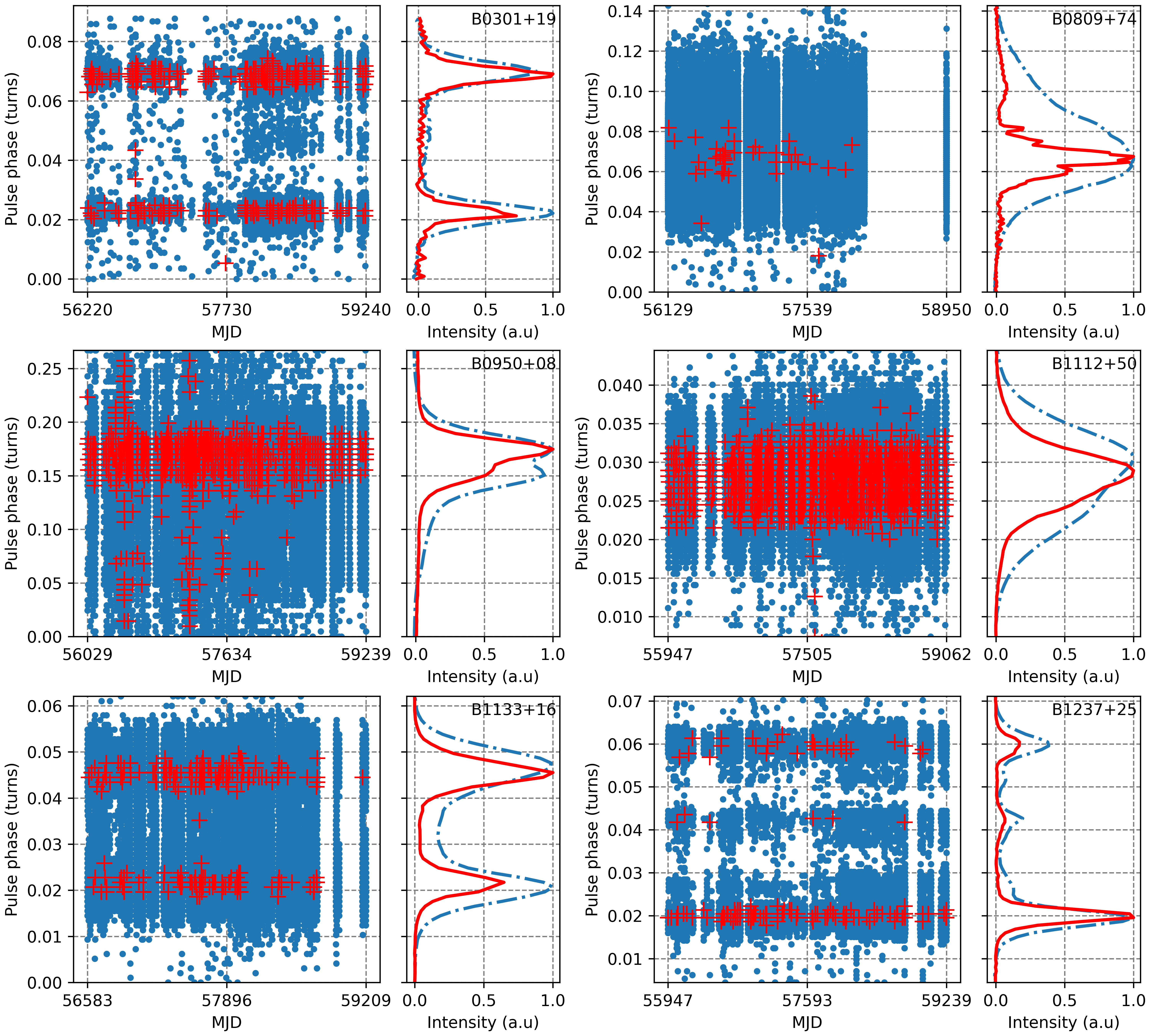}
\end{center}
\end{multicols}
\caption{Left panels: Phase distribution of non-giant (blue points) and giant (red points) pulses from the studied pulsars. {Right panels:} the averaged GRP (red solid line) and the non-giant averaged pulse (blue dashed line) studied pulsars}
\label{fig:phases}
\end{figure*}

The simple gaussian-like shape of an average pulse is typical for the pulsar B1112+50. However, there is an asymmetry of the averaged pulse shape (see right panel of Fig.~\ref{fig:phases}). GRPs from B1112+50 were distributed randomly along the average pulse longitude, which was narrower than the distribution of non-giant pulses. The averaged GRP was about two times narrower than the averaged pulse, which included all non-giant pulses from this pulsar. There was a significant offset between the peaks of the averaged GRP and average pulse. However, according to the fitting, the profile's location is quite similar (see Fig.~\ref{fig:fitting}). For averaged GRP, the mean of Gaussian form is equal to (46.76~$\pm$~0.06)~ms, for the average non-giant profile it is equal to (47.38~$\pm$~0.15)~ms. In units of the pulsar period, these locations are 0.0282 and 0.0286, respectively.

\subsection{PSR B1133+16}

In the analyzed 1050 observation sessions of the pulsar, 107~379 individual pulses with a signal-to-noise ratio of more than 4$\sigma$ were detected. Out of these, 236 individual pulses were classified as GRPs. The rate of GRPs generation of B1133+16 is visibly unstable (see Fig.~\ref{fig:rate_gp}, B1133+16). The maximum of the rate generation is 18 GRPs per hour. On average, the pulsar generates around four GRPs per hour. But for the B0809+76, there were no detections in numerous sessions. B1133+16 generates only one cluster of GRPs, shown in Fig.~\ref{fig:group_gp}.

B1133+16 has a double-component average pulse (see right panel of Fig.~\ref{fig:phases}). The distribution of non-giant pulses shows that these pulses could be recorded between the main components of the profile. In the great majority of cases, GRPs were detected on the longitudes of main components  only. The 60\% of registered GRPs were detected on the second component of the pulsar average profile.
The fitting results have shown that the distance between the main components for the averaged GRP and the average pulse is significantly different (see Fig.~\ref{fig:fitting}, B1133+16). For the average pulse, the distance is equal to (30.7~$\pm$~0.3)~ms and for averaged GRP the difference is equal to (28.49~$\pm$~0.12)~ms. In units of the pulsar period, these distances are 0.0259 and 0.0240, respectively.

\subsection{PSR B1237+25}

\begin{figure*}
\begin{multicols}{2}
\begin{center}
\includegraphics[width=2\columnwidth]{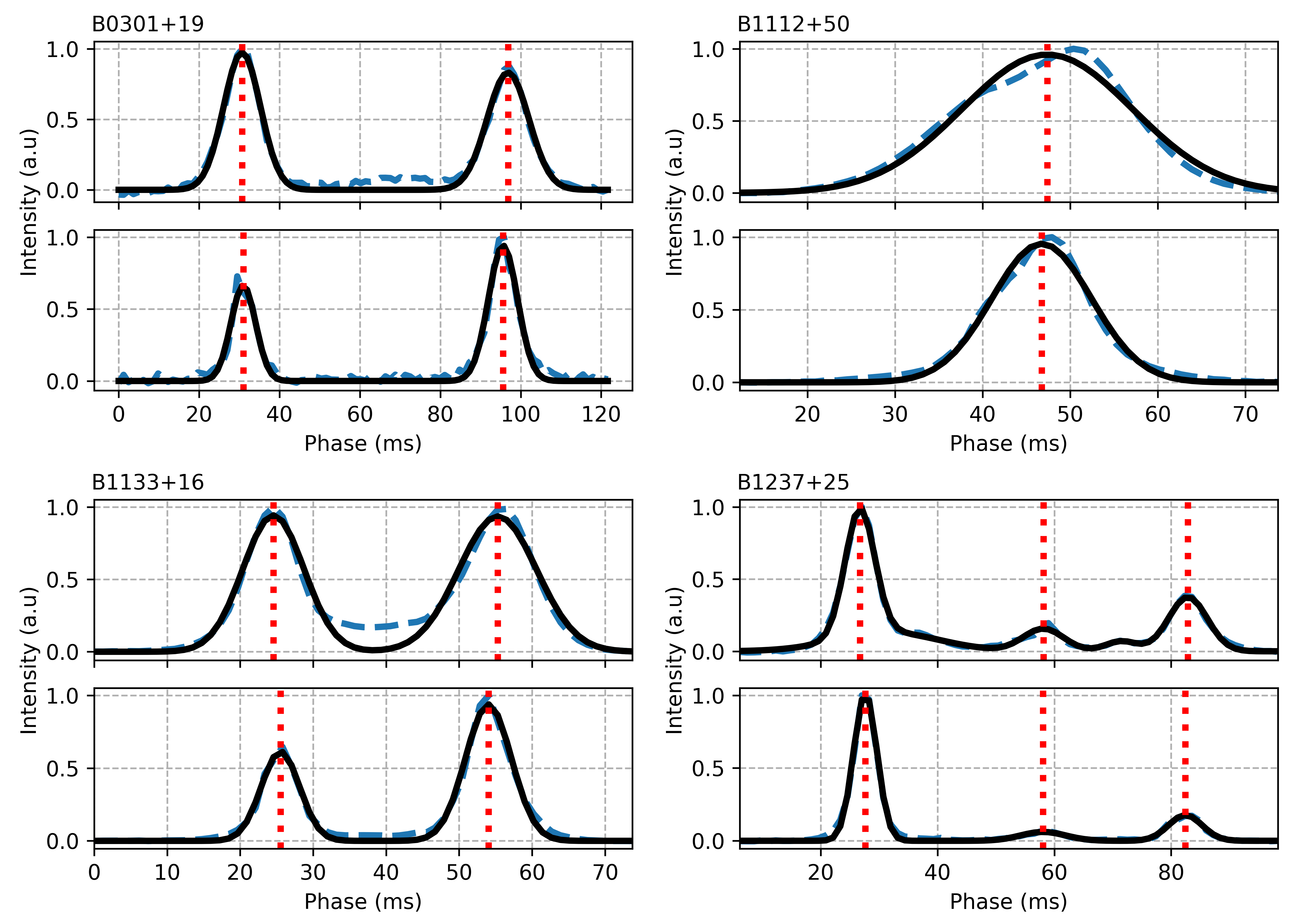}
\end{center}
\end{multicols}
\caption{Results of fitting of averaged pulses (upper panels) and the averaged GRPs (lower panels) for several studied pulsars.  Blue dashed lines show an averaged pulse and an averaged GRPs for each pulsar. Black solid lines show a corresponding fitting by combination of Gaussian function. Red doted lines show a location of a peak for the corresponding Gaussian function.}
\label{fig:fitting}
\end{figure*}

We analyzed 37~483 individual pulses from B1237+25 with a signal-to-noise ratio of more than 4$\sigma$. Out of these, 145  were classified as GRPs. The pulsar B1237+25 shows a quite stable GRPs generation rate (see Fig.~\ref{fig:rate_gp}, B1237+25). That were detected in 70\% of the observed sessions. On average, PSR B1237+25 generates approximately four GRPs per an hour. The maximum value of the GRPs rate is 34 giant pulses per hour. We detected only one cluster of GRPs (see it in Fig.~\ref{fig:group_gp}, B1237+25).
The average pulse of B1237+25 is very complex and includes five components. Only three components (1, 3, and 5) can be reliably identified in one observation session at low frequency. The sum of numerous individual pulses allowed us to detect all five components of the average pulse. As seen in Fig.~\ref{fig:phases}, the GRPs are emitted on the longitudes of the 1, 3, and 5 components. In 77 percent of cases, giant pulses are recorded in the first component of the average profile. The fifth component accounted for 19\% of detected giant pulses, the third - 4\%. We used the sum of five Gaussian functions (see upper panel of Fig.~\ref{fig:fitting}) for fitting the complex average pulse of B1237+25. As a result, the positions of 1, 3, and 5 components for average pulse are equal to (26.70~$\pm$~0.04)~ms, (58.4~$\pm$~0.20)~ms, and (82.94~$\pm$~0.09)~ms, respectively. The shape of averaged GRP is three components only, and we used the sum of three Gaussian functions to fit this profile (see lower panel of Fig.~\ref{fig:fitting}). The locations of the components of average GRP are (27.62~$\pm$~0.02)~ms, (58.1~$\pm$~0.5)~ms and (82.44~$\pm$~0.14)~ms. In this case, the difference between the most distant components  (1 and 5) for the average GRP and the average non-giant pulse is equal to (54.82~$\pm$~0.16)~ms and (56.24~$\pm$~0.12)~ms, respectively. In units of the pulsar period, these distances are 0.0397 and 0.0407.

\section{Discussion}
\label{sec:dis} 

The pulsars considered in the analysis (except B0809+74) are GRP emitters having small values of a magnetic field on a light cylinder. Existing theoretical models of GRPs generation \citep[e.g.,][]{Hankins2003, Istomin2004, Petrova2006a, Machabeli2019} have been basically aimed at explaining the power-law of the distribution of bright pulses and the extremely short duration of the GRPs, found in the pulsar B0531+21 \citep{Hankins2003}. Such parameters of the GRPs emission as generation rate or the duration of the period of activity have not been considered in these works. However, it seems to us very reasonable to hypothesize that since all the mentioned models are based on the basic pulsar parameters (the period of rotation and its first derivative). According to these models, the GRPs generation should be similar for pulsars with similar periods and period derivatives. Not only the pulse distribution, polarization, and duration of the GRPs should be similar, but also such parameters as the rate of generation of GRPs and the generation of clusters of GRPs. We have obtained the opposite result. Pulsars with highly similar physical parameters (e.g. B1112+50, B1133+16 and B1237+25) differ in rate of generation of GRPs and generation of clusters of GRPs.

According to our observations and analysis, the rate of generation of GRPs from studied pulsars varies significantly on the considered time interval. Each pulsar demonstrates at least one significant increase in the GRPs generation rate (see Fig.~\ref{fig:rate_gp}).

One can interpret this finding in several ways. First, it may reflect the instability of mechanism of GRPs generation. However, the emission rate of typical individual pulses is also subject to strong variability. In this way, the detected rate of GRPs may reflect the common nature of pulsed emission, both giant and normal.
 
Rates of generation of pulsars B1112+50 and B0950+08 are peculiar because they keep sufficiently high level during studied periods. We detected the largest number of clusters of GRPs from these pulsars in comparison with other ones. The difference cannot be explained by different durations of the observations: although the total time of observation of the pulsar B1112+50 is $\sim$3 times larger than the total time of observation of pulsars B1133+16 and B1237+25, the number of detected clusters being 75 times larger.

Another striking feature of B1112+50 is the longest sequence of GRPs. At period 1.6564~s (see Tab.~\ref{tab:pulsar_info}) duration of emission of this group of giant pulses is approximately seven seconds. It could mean that the process giving rise to GRPs in the magnetosphere of this pulsar is not burst-like, but extended in time. It is the only pulsar in our study which demonstrated such an extraordinary behavior.

The rates of GRPs generation of the pulsars B1237+25 and B1133+16 are similar -- they have almost the same value of the average rate: 0.06 and 0.07 $min^{-1}$, respectively. 

The analysis of phase distribution shows that GRPs are located on the longitudes of the main component of average profiles of pulsars. However, the variance in longitude for the GRPs is much less. Moreover, for pulsars with multicomponent average profiles, the distance between the components of the averaged profile produced using GRPs only is less than the same distance for the average profile that includes non-giant pulses. For B1133+16, this deference between distances mentioned above is (2.2~$\pm$~0.4)~ms. Taking into account the model of nested cone structure \citep{rankin1993, gil1993}, this difference, as well as the narrow distribution of the GRPs over the longitudes of the components, quite likely indicates that the GRPs of these pulsars are generated closer to the surface of the neutron star. Analysis of the phase distribution of giant pulses of the pulsar B0531+21\citep{Bij2021} has not shown a noticeable difference in the arrival times of strong and weak pulses.

Detected sinusoidal variations in the rate of B0950+08 GRPs cannot be associated with any distortions in the telescope (change in the effective area or change in the gain), since the GRP sampling criterion we used takes into account factors associated with the telescope. Something similar is also observed from B1133+16. This behavior of the generation rate can be most interesting when applied to the formation of GRPs as an energy accumulation mechanism in pulsar magnetosphere plasma \citep{Machabeli2019}. At the same time, the study of the rate of generation from MJD 57638 to MJD 58624 \citep{kuiack2020} has given estimates of 30 GRPs per hour, while our estimates indicate this value as the average value of the rate of generation for B0950+08. 

In addition to the fact that the average profile of the pulsar B0809+74, which includes only the GRPs, strongly differs from the average profile of the non-giant pulses. On the distribution of the GRPs phase (right panel of Fig.~\ref{fig:phases}, B0809+74), one can notice a clear drift in the position of the GRPs along the longitudes of the average profile. It is known that this pulsar has a pronounced drift of subpulses, first discovered by \cite{page1973} and further studied in detail \citep[e.g.,][]{backer1975, rankin2006, rankin2014}. Due to the lack of observations of this pulsar in 2018 and 2019, we cannot state with confidence that the systematic drift of the GRPs generation region in the phase of the average profile is not accidental. Further observations of this pulsar will help to confirm or refute the feature we discovered.

\section{Conclusions}
\label{sec:conclusions} 

The analysis of variations in the rate of generation of powerful individual pulses of six pulsars of the Northern Hemisphere at 111 MHz on a time scale of nine years is presented. 

It is shown that pulsars B0301+19, B1112+50, B1133+16, and B1237+25, which have been noted as GRPs generators and have similar physical parameters (rotation period, period derivative, and the magnetic field on the light cylinder), demonstrate different generation rates  and clusters generation. Each of the studied pulsars demonstrate jumps in the rate of GRPs generation by 2-4 times from the average level. In the pulsar B1237+25, a maximal generation rate was found to exceed 10 times the average level. In the overwhelming majority of observations, the value of the generation rate is below the corresponding upper control limit.

GRPs from these pulsars are distributed along the longitudes of the main components of an average pulse. This distribution is 1.5-2 times narrower than the phase distribution of non-giant pulses. A systematic shift in the position of the GRPs was found in the pulsar B0809+74. 

For pulsars with multicomponent average profiles, it is found that the distance between the components of the average GRP profile and the distance between the components of the average non-giant profile differ substantially. For pulsars B0301+19 and B1237+25, this difference is about 1.5~ms. For pulsar B1133+16, this difference is equal to 2.24~ms.

It is shown that the pulsar B1112+50 more often than others generates GRPs clusters. Besides, this pulsar demonstrates the longest sequence of GRPs containing four single pulses. Pulsar B0950+08 also demonstrates a significant number of a pair of GRPs.

\section*{Acknowledgements}
\label{sec:acknowledgements} 

The authors are grateful to the program committee of the PRAO ASC LPI for the allocated observing time. AK acknowledges the support by the Foundation for the Advancement of Theoretical Physics and Mathematics <<BASIS>> grant 18-1-2-51-1. We thank our colleagues E.~V.~ Kravchenko, V.~A.~Potapov and M.~S.~Pshirkov for useful discussions and contributions during the preparation of this paper.


\section*{Data availability}
The data underlying this article will be shared on reasonable request to the corresponding author.


\bibliographystyle{mnras}
\bibliography{main}
\newpage





\label{lastpage}
\end{document}